\documentclass[prb,reprint,twocolumn,superscriptaddress,noshowpacs,notitlepage,longbibliography,10pt,citeautoscript]{revtex4-2}%

\usepackage{pdfpages} 
\usepackage{pgffor} 

\makeatletter
\AtBeginDocument{\let\LS@rot\@undefined}
\makeatother

\def\supplementfilename{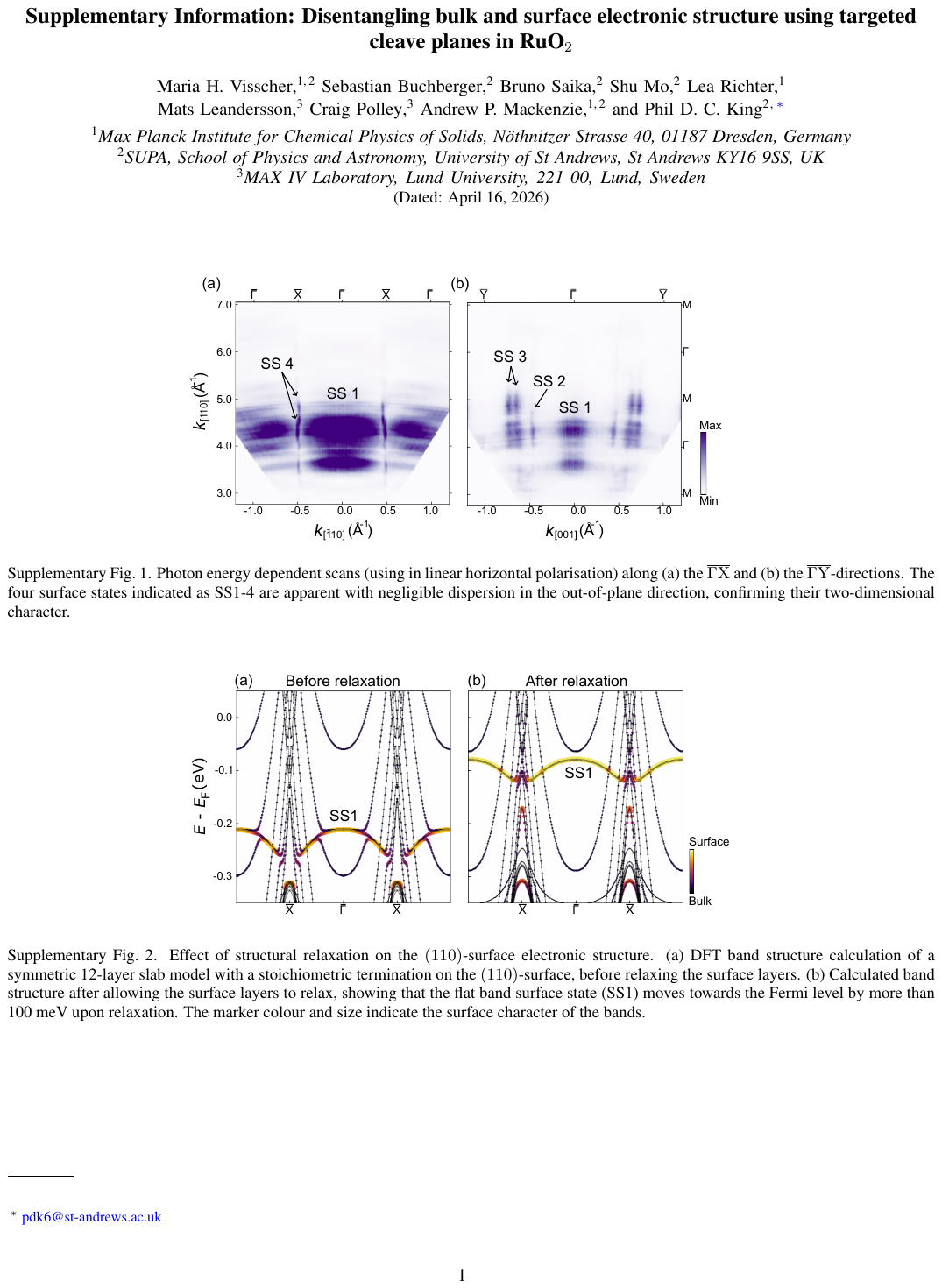}

\pdfximage{\supplementfilename}
\def\numbersupplementpages{\the\pdflastximagepages}

\newif\ifarXiv
\arXivtrue

\usepackage{graphicx,bm,times}
\graphicspath{ {./figures/main/} {./figures/supp/} }
\usepackage{amsmath}
\usepackage{amsfonts}
\usepackage{amssymb}
\usepackage{mathtools}
\usepackage{color}
\usepackage{hyperref}
\usepackage[version=4]{mhchem}
\usepackage{gensymb}
\hypersetup{
  colorlinks = true,
  allcolors = {blue}
}
\usepackage[colorinlistoftodos]{todonotes}
\begin{document}

\title{Disentangling bulk and surface electronic structure using targeted cleave planes in RuO$_2$}

\author{Maria~H.~Visscher}
\affiliation{Max Planck Institute for Chemical Physics of Solids, N{\"o}thnitzer Strasse 40, 01187 Dresden, Germany}
\affiliation{SUPA, School of Physics and Astronomy, University of St Andrews, St Andrews KY16 9SS, UK}

\author{Sebastian~Buchberger}
\author{Bruno~Saika}
\author{Shu~Mo}
\affiliation{SUPA, School of Physics and Astronomy, University of St Andrews, St Andrews KY16 9SS, UK}

\author{Lea~Richter}
\affiliation{Max Planck Institute for Chemical Physics of Solids, N{\"o}thnitzer Strasse 40, 01187 Dresden, Germany}

\author{Mats~Leandersson}
\author{Craig~Polley}
\affiliation{MAX IV Laboratory, Lund University, 221 00, Lund, Sweden}

\author{Andrew~P.~Mackenzie}
\affiliation{Max Planck Institute for Chemical Physics of Solids, N{\"o}thnitzer Strasse 40, 01187 Dresden, Germany}
\affiliation{SUPA, School of Physics and Astronomy, University of St Andrews, St Andrews KY16 9SS, UK}

\author{Phil~D.~C.~King}\email{pdk6@st-andrews.ac.uk}
\affiliation{SUPA, School of Physics and Astronomy, University of St Andrews, St Andrews KY16 9SS, UK}

\date{\today}
\begin{abstract}
  Rutile RuO$_2$ has attracted significant interest due to its putative unconventional electronic and magnetic properties and its proximity to superconductivity.
  However, the measurement and interpretation of its electronic structure has been complicated by a strongly three-dimensional crystal structure.
  Here, we demonstrate how the preparation of targeted $(110)$ and $(100)$ surfaces via focused ion beam (FIB)-engineered cleaving allows the acquisition of high-quality measurements of the electronic structure using angle-resolved photoemission spectroscopy.
  Our results demonstrate that ARPES spectra of RuO$_2$ are, in fact, largely dominated by signatures of distinct surface electronic states.
  From comparison with density-functional theory, we resolve a surface termination-dependent variation of these, and disentangle them from highly-three-dimensional bulk states and surface resonances.
  Moreover, we find a marked role of the substantial spin-orbit coupling of the Ru 4$d$ orbitals in the surface region, where a breaking of spatial inversion symmetry leads to significant Rashba-type spin splittings of the surface bands.
\end{abstract}

\maketitle

\section*{Introduction}

Ruthenium dioxide is a rutile-structured oxide metal, well-known for its excellent catalytic activity \cite{Over:2012}.
Recently, it has also come to prominence for its electronic and potential magnetic properties: it hosts superconductivity under strain~\cite{Ruf:2021, wadehra_strain-induced_2025} while Dirac nodal lines are found in its electronic structure as a result of its non-symmorphic crystal structure~\cite{Sun:2017}.
The presence of such symmetries has also led to suggestions that RuO$_2$ could host an altermagnetic phase.~\cite{Smejkal:2022, Gonzalez:2021, Feng:2022, Tschirner:2023, Bai:2023, Karube:2022}.
While recent investigations indicate that the bulk material is, in fact, non-magnetic \cite{Hiraishi:2024, Kessler:2024, Kiefer:2025}, it has theoretically been predicted to host surface magnetism~\cite{Ho:2025, Torun:2013}.
Indeed, given the three-dimensional crystal structure, a distinct surface magnetic and electronic environment can naturally be expected.
Quasi-1D ``flat-band'' surface states have already been observed~\cite{Jovic:2018}, which have been proposed to support RuO$_2$'s catalytic activity~\cite{Jovic:2021}.
Such a distinct surface environment may also provide a natural route to understand reported spin-polarised signatures in angle-resolved photoemission (ARPES) measurement \cite{Liu:2024}.

Together, this motivates a re-examination of the bulk vs.\ surface electronic structure in RuO$_2$.
ARPES should be an ideal probe for this~\cite{Jovic:2018, Liu:2024, Osumi:2026, Lin:2025}.
Such studies are, however, significantly complicated by the strongly three-dimensional crystal structure.
As a result, the material lacks a natural cleaving plane, which hinders the preparation of the required flat and clean surfaces by conventional methods.
Secondly, significant $k_z$-broadening can be expected, which can add substantial challenges for identifying the relevant bulk states, especially when paired with limited effective angular resolution stemming from rough cleave surfaces.
Indeed, although quantum oscillations measurements show good agreement with calculations of the bulk electronic structure from density functional theory \cite{Graebner:1976, Yavorsky:1996, Wu:2025}, ARPES studies have proposed bulk bands which, in some cases, require significant energetic shifts to match those in the calculations~\cite{Jovic:2018, Liu:2024, Lin:2025}.

To investigate the origin of these apparent discrepancies and to overcome the challenges associated with the three-dimensional crystal structure, here we utilise a fabrication method~\cite{Hunter:2024} based on Focused Ion Beam (FIB) structuring to stimulate cleaving along desired crystallographic planes.
This allows us to obtain high quality surfaces oriented along the $(110)$- and $(100)$-directions, making high resolution ARPES experiments possible.
From these experiments and supporting bulk and surface electronic structure calculations, we identify a rich spectrum of bulk and surface bands, the latter of which exhibit a strong dependence on surface stoichiometry and structural relaxations.
We further identify a marked influence of spin-orbit coupling on the surface electronic states, giving rise to significant spin-splittings in the broken inversion symmetry environment that the surface hosts.

\section*{Results}
\subsection*{FIB designed cleaving planes in ruthenium dioxide}
\begin{figure*}
  \centering
  \includegraphics[width=\textwidth]{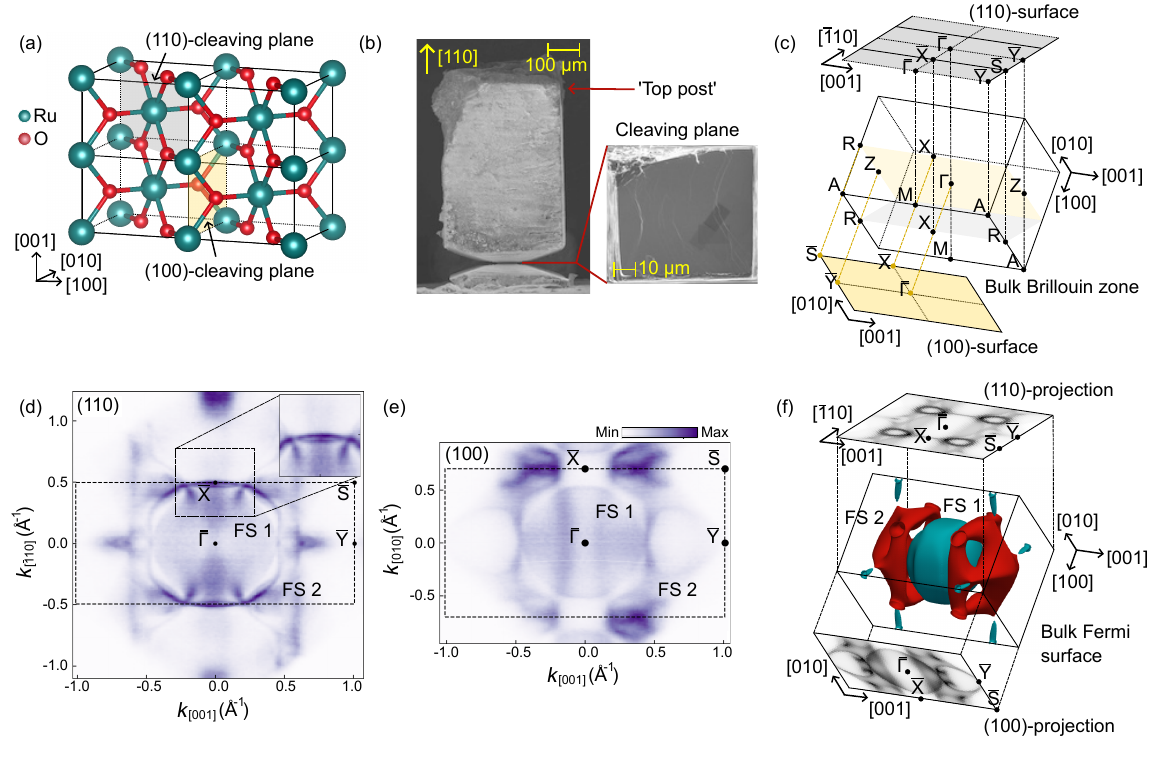}
  \caption{Fermi surfaces from FIB designed cleaving planes in ruthenium dioxide.
    (a) Crystal structure of $\textrm{RuO}_2$.
    The $(110)$- and $(100)$-cleaving planes are indicated in grey and yellow.
    (b) Scanning Electron Micrograph of a FIB-structured sample, which allows for designed cleaving planes along the $(110)$ (as shown) and $(100)$-surface.
    The cleaved surface is shown inset.
    (c) Schematic of the bulk Brillouin zone, and its projections onto the $(110)$ and $(100)$ surfaces.
    (d), (e) ARPES Fermi surfaces of the $(110)$ and $(100)$-surfaces, measured at $h\nu=68$~eV using linear vertical light polarisation.
    (f) The calculated bulk Fermi surface and its projections onto the $(110)$- and $(100)$ surface Brillouin zone to simulate the expected intensity in the ARPES measurements shown in (d,e).
    The grey and yellow highlighted planes in the bulk zone indicate the XR- and $\Gamma$XRZ-planes.
  }
  \label{fig:fig1}
\end{figure*}
Ruthenium dioxide crystallises in the tetragonal rutile crystal structure (Fig.~\ref{fig:fig1}a), comprised of edge-sharing RuO$_6$ octahedra.
This renders the structure highly three-dimensional, lacking a good natural cleaving plane.
As a result, when conventional top-post cleaving methods are used to prepare clean surfaces for ARPES, they typically result in small exposed facets oriented along the $(110)$-direction, while high-quality surfaces of other orientations are extremely difficult to obtain~\cite{Osumi:2026}.
To improve the cleaving of this compound, we utilise here a method pioneered by Hunter et al.~\cite{Hunter:2024}, where focused ion beam (FIB) structuring is used to stimulate cleaving along a desired crystalline direction.
An example of this is shown in Fig.~\ref{fig:fig1}b.
The sample is first cut from a bulk crystal to form a pillar, after which a FIB is used to create a narrow constriction close to the base of the sample.
The final cleaving plane is then defined by milling a thin line across the constriction.
This notch acts as a `strain lens' that focuses the stress induced during cleaving into one weak point.
The upper half of the sample acts as the `top post' of a conventional cleaving setup, so that when we apply force to it, the sample cleaves along this defined plane with a smooth and flat surface (see inset in Fig.~\ref{fig:fig1}b).
Using this approach, we were able to obtain facets of at least $20\times20$~$\mu\rm{m}^2$ for the $(110)$-orientation, and $10\times10$~$\mu\rm{m}^2$ for the less favourable $(100)$-orientation.
Such surfaces are well-suited for ARPES experiments using small ($\lesssim{10}$~$\mu$m) photon beams (see Methods).

\begin{figure*}
  \centering
  \includegraphics[width=\textwidth]{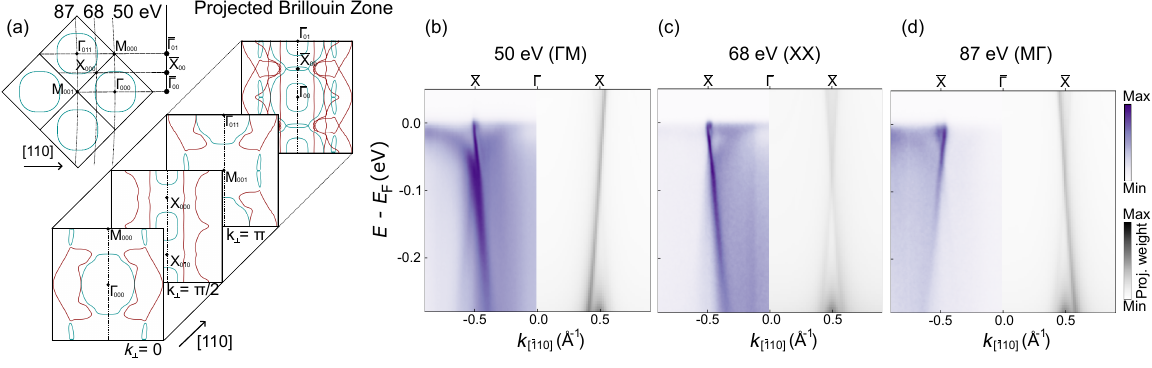}
  \caption[width=\textwidth]{
    Reconciling apparent bulk band crossing from surface projections.
    (a) Schematic of how the bulk Brillouin zones stack together and project onto the surface Brillouin zone.
    The coloured lines show the bulk Fermi surface contours extracted from DFT.
    (b--d) Dispersions along $\overline{\Gamma \textrm{X}}$ at photon energies of 50~eV, 68~eV, and 87~eV.
    These probe planes centered on $\Gamma$MAZ, XR, and MA$\Gamma$Z, respectively.
    Left shows the ARPES data, compared with a bulk projected DFT calculation (including $k_\perp$-integration) on the right.
    The color bars indicate the photoemission intensity and the simulated projected weight, respectively.
  }
  \label{fig:fig2}
\end{figure*}
\subsection*{Fermi surfaces and bulk nodal lines of RuO$_2(110)$ and RuO$_2(100)$}
We show the measured Fermi surfaces from samples prepared in this way in Fig.~\ref{fig:fig1}d,e.
These were taken in measurement conditions (i.e., at a photon energy of 68 eV) which should probe the region centered around the XR- and $
\Gamma $XRZ planes for the $(110)$- and $(100)$-orientation, respectively.
However, due to the surface sensitivity of ARPES, we expect significant broadening of our measured features along the out-of-plane ($k_\perp$) direction ($[110]$ or $[100]$, respectively), contributing to the measured spectra.
The measured Fermi surfaces are clearly strongly affected by this $k_\perp$-broadening.
Despite exhibiting some sharp and clearly-dispersive features (pointing to the good sample cleave quality), the majority of the surface Brillouin zone is filled with broad spectral weight for both surface orientations.
From inspection of our calculated Fermi surface from DFT (Fig.~\ref{fig:fig1}f), it is clear that the central, almost circular, filled-in spectral weight for both the $(110)$ and $(100)$ surfaces originates from the near-spherical Fermi surface, centred at $\Gamma$ (FS1).
Meanwhile, the FS2 pockets contribute spectral weight shifted away from the $\Gamma$ pocket along $k_{[001]}$.
Both Fermi surfaces disperse strongly along all three momentum directions, and as a result, they are substantially broadened in our measurements.
To simulate the effect of this, we project our calculations onto the $(110)$- and $(100)$-surfaces, accounting for the $k_\perp$-broadening expected to be present in the experimental measurements (Fig.~\ref{fig:fig1}f, see Methods).
While the intensity of some features are overestimated (e.g. the sharp rings offset from the $\overline{\rm{X}}$ points, likely reflecting matrix element variations not captured in this projection), our calculations well reproduce many of the spectral features we observe, including essentially all of the filled-in spectral weight.

Of particular interest are the features near the $\overline{\rm{X}}$-point of the $(110)$-surface, visible in the inset in Fig.~\ref{fig:fig1}d.
Here, Fermi pockets centred at $\overline{\Gamma}_{00}$ and the neighbouring $\overline{\Gamma}_{01}$ appear to cross, leading to arcs forming at the $\overline{\rm{X}}$-point.
Similar band crossings can also be seen in our measured dispersions along the $\overline{\Gamma\rm{X}}$ direction (Fig.~\ref{fig:fig2}b-d), and were previously attributed to one of the material's Dirac Nodal lines \cite{Jovic:2018}.
Such a nodal crossing is present in DFT calculations, but is not expected to intersect the Fermi level.
Indeed, to reproduce the measured spectra required shifting the DFT bands downwards by $\approx600$~meV, seemingly at odds with the good agreement between the DFT Fermi surface and quantum oscillations~\cite{Graebner:1976, Wu:2025}.

We attribute this discrepancy to how the bulk features project onto the surface Brillouin zone due to the strong $k_\perp$-integration in ARPES.
Fig.~\ref{fig:fig2}a illustrates how the three-dimensional Brillouin zones stack along $k_{[110]}$.
Because of this stacking, the periodicity of the surface Brillouin zone is halved along $k_{[\overline{1}10]}$ compared to the bulk zone.
The extremal edge of FS1 along the $\Gamma_{000}\rm{M}_{000}$-directions extends beyond the edge of the surface Brillouin zone and projects partly onto the next zone.
At the surface zone boundary, it then overlaps with the projection of FS1 that originates in the second bulk Brillouin zone (along $\rm{M}_{001}\Gamma_{011}$) but extends slightly into the first zone.
This gives rise to the appearance of a ``crossing'' of these two Fermi surfaces and the arcs around the $\overline{\rm{X}}$-points, illustrated in Fig.~\ref{fig:fig2}a and visible in the Fermi surface of Fig.~\ref{fig:fig1}d.
Since the apparent crossing is the result only of a projection, the intensity of its two branches changes with photon energy, as the central $k_\perp$ value probed is tuned.
This is visible in our measured dispersions and simulated ARPES projections (Fig.~\ref{fig:fig2}b-d).
At a photon energy of 50 eV (probing dominantly the region centred on $\Gamma_{000}\rm{M}_{000}$), the positive dispersion contributes most strongly (additional sharper surface-derived features are also visible in the experiment which will be discussed further below).
This spectral weight shifts to the negative dispersion at 87 eV (probing the region centred on $\rm{M}_{001}\Gamma_{011}$).
At 68 eV, probing the region centred on the XX-line, both branches contribute similar spectral weight, leading to the apparent crossing and the arcs observed in Fig.~\ref{fig:fig1}d.
It is these crossing features that were previously assigned as a Dirac Nodal line.
We conclude, however, that this is in fact a projection artefact of the measurement.
Importantly, the measured band features, including this crossing in the surface projection, are well accounted for in our simulations from bulk DFT calculations.
We thus conclude that no substantial energy shift of the DFT bands is required to match the experimentally observed dispersions.

\begin{figure*}
  \centering
  \includegraphics[width=\textwidth]{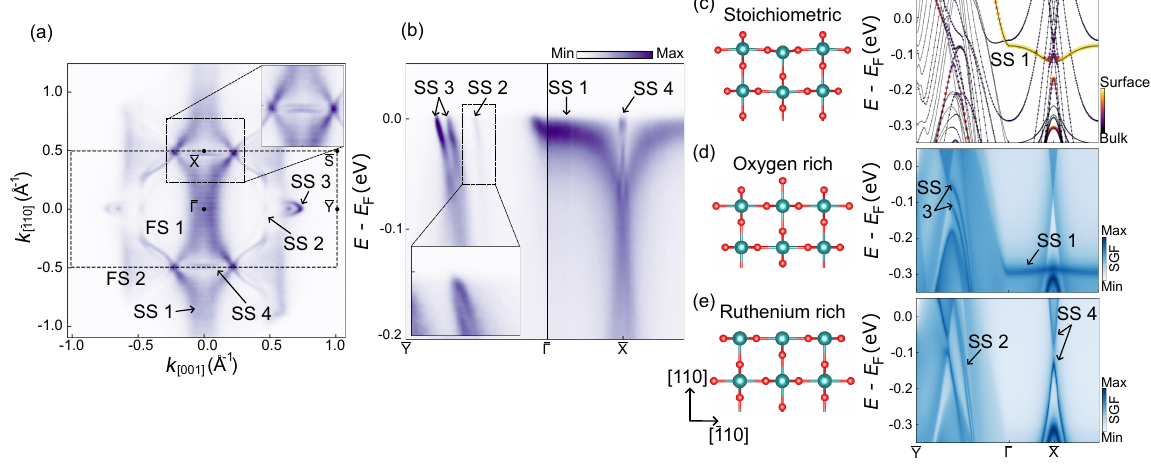}
  \caption{
    Termination-dependent surface resonances on the $(110)$-surface.
    (a) Measured $(110)$-Fermi surface ($h\nu=68$~eV, linear horizontal light polarisation), highlighting the surface resonances SS1-SS4.
    (b) Corresponding dispersions measured along the $\overline{\rm{Y} \Gamma \rm{X}}$ directions, showing the four surface resonances.
    The inset shows SS2 measured in linear vertical light polarisation, highlighting a small band splitting.
    (c) Surface structure of the stoichiometric termination of the $(110)$ surface, and the corresponding surface electronic structure from DFT slab calculations.
    The marker colour and size indicate the projected surface weight.
    (d) Surface structure of the oxygen rich termination, and corresponding surface Green's Function calculation, showing SS1 and SS3.
    (e) Surface structure of the ruthenium rich termination, and its surface Green's Function calculation, showing SS2 and SS4.
  }

  \label{fig:fig3}
\end{figure*}

\subsection*{Surface states}

Nonetheless, besides these projected bulk bands (which contribute rather broad spectral weight in Fig.~\ref{fig:fig1}d,e and Fig.~\ref{fig:fig2}b-d), additional sharper states are visible in our measurements which are not captured by our bulk DFT calculations.
These additional states are most clearly visible in linear horizontal light polarisation, and we show measurements of the corresponding Fermi surface and dispersions from the $(110)$ surface in Fig.~\ref{fig:fig3}a,b.
A bright but somewhat broad quasi-1D Fermi surface is visible as a near-vertical stripe in our measured Fermi surface.
This derives from the flat band visible along $\overline{\Gamma}\overline{\rm{X}}$ in our dispersion (Fig.~\ref{fig:fig3}b, marked SS1).
We attribute this as the flat band surface state observed previously~\cite{Jovic:2018}, which was found to originate from ruthenium-oxide chains on the $(110)$-surface~\cite{Jovic:2021}.
While this appears qualitatively similar to its observation in previous studies, we note that we can resolve here an additional, previously unobserved, splitting in the band in the vicinity of the $\overline{X}$ point (see also figure \ref{fig:fig2}c).
This may suggest a significant hybridisation between the bulk and surface bands here.
This, as well as the generally-broad nature of this state, is consistent with the fact that the surface state disperses through the momentum space region onto which the highly-dispersive bulk bands project.
It should therefore rather be considered as a surface resonance than an isolated surface state.

In addition to this flat band, we also observe several much more dispersive states.
SS2 has been identified as a surface state by Liu~{\it et al.}~\cite{Liu:2024}, while the pair of states along $\overline{\Gamma\rm{Y}}$ (which we attribute as SS3) were observed previously but assigned as bulk states~\cite{Jovic:2018, Liu:2024}.
There is, however, a significant energy mismatch between this feature and the calculated bulk band.
As discussed, once surface projection effects are accounted for, we find a good general agreement between calculation and observed bulk features.
Moreover, we find that the states we denote as SS3 have a dispersion which is independent of the probing photon energy (Supplementary Fig.~1).
We thus also attribute this state as a surface state/resonance here.
Finally, an additional sharp state (SS4) is visible close to the $\overline{X}$ point of the surface Brillouin zone, contributing additional sharp and $h\nu$-independent bands in Fig.~\ref{fig:fig2}b-d.

To better understand the origin and nature of this rich array of apparently surface-related bands, we have performed surface electronic structure calculations with DFT.
We show in Fig.~\ref{fig:fig3}c calculations (see Methods) modelling the surface with a symmetric 12-layer slab with a stoichiometric termination on both sides.\
Consistent with prior work \cite{Jovic:2018, Jovic:2021, Ho:2025}, these calculations reproduce the flat band surface resonance (SS1), although they locate it at a slightly higher binding energy than observed in our experimental measurements.
We note that the binding energy of this state appears highly sensitive to surface relaxation in our DFT calculations (Supplementary Fig.~2), again likely reflecting a strong coupling of the surface band to the underlying bulk.
Our DFT surface projections also reveal some increased wavefunction weight at the surface for states close to the $\overline{X}$ point and between the $\overline{\Gamma}\overline{\rm{Y}}$ direction.
However, clearly resolving these as distinct surface bands from the strongly-dispersive bulk states is challenging for feasible sizes of a DFT supercell.
To better capture these, we have therefore performed additional surface state calculations using a surface Green's function (SGF) method (see Methods).
These model a surface on a semi-infinite system, where we consider bulk-like truncations of the unit cell.
While this does not capture the effects of surface relaxations, it does allow us to model the effect of having an oxygen rich surface vs.\ a ruthenium rich one (Fig.~\ref{fig:fig3}d,e): both terminations that could be expected experimentally \cite{Over:2002}.
Due to a lower computational cost, they also allow us to include the effects of spin-orbit coupling (SOC) in our calculations (see Supplementary Fig.~3 for a comparison with and without SOC).

The calculated spectra for the oxygen and ruthenium rich surfaces are shown in figure \ref{fig:fig3}d-e.
Together, these show features that are in excellent qualitative agreement with our measured band structures.
The oxygen-rich calculation reproduces the `flat-band' surface state between the $\overline{\Gamma}$ and $\overline{X}$ directions, albeit located at a significantly too high binding energy, which is consistent with previous work \cite{Liu:2024, Osumi:2026}.
We attribute the binding energy overestimation to the effects of structural relaxation at the surface, which we find in our DFT slabs to dramatically lower the energy of this flat band surface state (Supplementary Fig.~2).
Interestingly in the experiment, we find that the flat band is located only $\approx10$~meV below the Fermi level, suggesting a modest surface doping or strain could move its associated large density of states to $E_\mathrm{F}$, potentially triggering a surface instability to a magnetic or other collective state.

Our calculations also reproduce a pair of additional surface resonances (marked SS3), located at the momentum space location at which the bulk Dirac nodal line (DNL2 from \cite{Sun:2017}) intersects the $(110)$-plane.
These are strongly split in the vicinity of this point, consistent with experimental observations from our measured Fermi surfaces and dispersions (Fig. \ref{fig:fig3}a, b).
Our calculations show how this splitting is a consequence of spin-orbit coupling, which we find lifts the spin degeneracy of these states (Supplementary Fig. 3, see also~\cite{Liu:2024}).
We therefore attribute the band splitting to a Rashba-like spin-orbit coupling in the presence of surface inversion symmetry breaking.

While the oxygen-rich calculations naturally reproduce SS1 and SS3, they do not capture SS2 and SS4.
These, however, can be realised in calculations of a Ru-rich surface.
SS2 is again split by spin-orbit coupling (Supplementary Fig.~3), albeit with a much smaller splitting than for the O-rich surface.
This splitting is, however, still resolvable with a small band splitting visible in our experimental measurements (inset Fig.~\ref{fig:fig3}b).
This points to the high surface quality resulting from our FIB-structured cleave planes.
While a good macroscopic cleave surface is obtained, micro-scale spatial variation (smaller than the size of our probing light spot) can still be expected in the surface termination, given the lack of a natural cleaving plane.
We will average over these in our measurements, explaining the simultaneous presence of surface states from distinct terminations in our measured ARPES data.

\begin{figure}
  \includegraphics[width = \columnwidth]{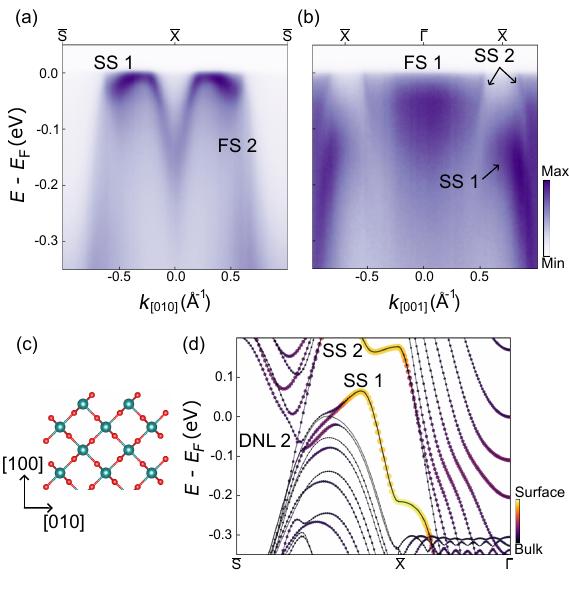}
  \caption{
    Surface resonances of the $(100)$-surface.
    (a), (b) Measured dispersions along the $\overline{\rm{XS}}$ and $\overline{\Gamma \rm{X}}$ directions ($h\nu=68$~eV, linear horizontal polarisation), showing bulk projected states coming from the two Fermi sheets (FS 1, 2) and two surface resonances (SS 1, 2).
    (c) Stoichiometric termination of the $(100)$-surface.
    (d) DFT slab calculation for this termination, showing the two surface states.
    The marker color and size indicates the projected surface weight.
  }
  \label{fig:fig4}
\end{figure}

As discussed above, FIB-structured cleaving also allows us to deterministically cleave along the distinct $(100)$ surface orientation.
We show the corresponding ARPES dispersions in Fig.~\ref{fig:fig4}a,b (see also Fermi surface measurements in Fig.~\ref{fig:fig1}e).
We observe broad spectral weight derived from the highly three-dimensional FS1 and FS2 bulk Fermi surfaces.
In addition, we observe a significant enhancement of spectral weight close to the bulk projected band gap along $\rm{\overline{XS}}$ and on the edge of the projection of the bulk FS1 states along $\rm{\overline{\Gamma X}}$.
This is again suggestive of additional surface-derived bands.
We denote the strong spectral weight along $\rm{\overline{XS}}$ as arising due to a surface state SS1, which has previously been assigned as having a drumhead-like character~\cite{Osumi:2026}.
In this respect, we note a significant dispersion of this state here, forming a V-shaped band around the $\rm\overline{X}$ point.
In addition, we observe a sharp band along $\rm{\overline{\Gamma{}X}}$ which we denote SS2.

These assignments are in good agreement with our DFT slab calculations performed for the $(100)$ surface.
For the stoichiometric slab (Fig.~\ref{fig:fig4}c), we already find two additional surface states (SS1 and SS2) whose dispersions are in good general agreement with our measured electronic structures (Fig.~\ref{fig:fig4}d).
The binding energy of these surface states is again strongly impacted by surface relaxations (Supplementary Fig.~4), with SS1 gaining additional dispersion and relaxing towards the bulk projected band gap.
This again points to a significant hybridisation between bulk and surface bands.
We note that for the $(100)$ surface, the stoichiometric surface appears to better describe the measured electronic structure than calculations for oxygen and ruthenium rich terminations (Supplementary Fig.~5).

\section*{Conclusions}
Together, therefore, our measurements and calculations point to a very rich surface environment in RuO$_2$.
A highly three-dimensional bulk couples to dangling-bond-derived states at the surface, providing multiple potential channels to enhanced catalytic activity.~\cite{Jovic:2021}
Substantial bulk-surface hybridisation is evident via the formation of surface resonances, while surface relaxation appears to favour surface bands pushed towards projected band gaps as much as possible.
The inversion symmetry breaking inherent to the surface -- together with a marked spin-orbit coupling of Ru $4d$ orbitals -- further allows significant Rashba-type spin splittings to develop.
This provides a potential route to rationalise previous spin-resolved signatures in surface-sensitive spectroscopic measurements of RuO$_2$.
More broadly, our measurements highlight that the measured spectra in this system (at least where dispersive band-like features are resolved) are dominated by surface rather than bulk states.
Significant care must therefore be taken in making altermagnetic assignments -- an inherently bulk property -- in this and related highly three-dimensional compounds, from ARPES alone.

{\small
  \section*{Methods}
  \noindent{\bf{Sample preparation}}
  Single crystals of \ce{RuO2} were grown by chemical vapour transport, using \ce{TeCl4} as transport agent~\cite{oppermann1975}.
  The amount of \ce{TeCl4} was calculated to yield a final pressure in the ampoule of approximately 1 bar at the target temperature.
  Powder of \ce{RuO2} (Alfa Aesar, 99.9\%, anhydrous) was mixed thoroughly with \ce{TeCl4} (Fisher Scientific, 99.9\%) under inert atmosphere.
  The final mixture was loaded into a quartz ampoule, evacuated and sealed.
  The ampoule was introduced into a horizontal two-zone furnace and heated for 10 days to 950 $\degree \rm{C}$ at the source and 820 $\degree \rm{C}$ at the sink.
  Finally, millimetre-sized, prism-shaped crystals were collected on the sink side of the ampoule.

  RuO$_2$ crystals were cut into pillars of $\approx 300\times300\times600$~$\mu$m$^3$ with their long axis aligned along a desired cleaving direction, here the $[110]$- or $[100]$-direction.
  FIB was then used to create a narrow constriction near the base of the sample.
  First, two triangular patterns were milled at high ion currents (0.2 -- 0.5 $\mu$A) and an acceleration voltage of 30 kV, reducing the width to $\approx 100 \ \mu$m.
  The constriction was then narrowed to 50 $\mu$m with two rectangular patterns at a lower ion current (60 nA).
  Afterwards, the pillar was rotated by 90 degrees to also reduce the cross-section from the orthogonal side.
  Again, first rough triangular patterns were cut at high ion currents (0.2 -- 0.5 $\mu$A) to reduce the width to $\approx 100 \ \mu$m.
  Afterwards, triangular patterns were milled, first at 15 nA to decrease the width to $\approx 80 \ \mu$m, and then at 4 nA to decrease the width to $\approx 60 \ \mu$m.
  This made it possible to use smaller ion currents to get sharper cuts, while having a reasonable process time.
  Lastly, a thin line was milled across the neck of the constriction at a current of 0.79 nA to define the final cleaving plane.
  The linecut was extended slightly beyond the neck, to ensure that its edges were also patterned and the constriction was fully defined from two perpendicular directions.
  For the $(110)$-orientation, a Xe-plasma FIB was used for all cuts.
  For the less favourable $(100)$-direction, a Ga-FIB was used for the final steps at low currents, as this was found to give smoother cuts.

  \

  \noindent{\bf{Angle-resolved Photoemission Spectroscopy}}
  Micro-ARPES measurements were performed at the Bloch A-branch end station of the MAX IV synchrotron with a beam spot size of $\approx 10 \ \mu$m.
  The samples were cleaved in ultra-high vacuum, using  a custom sample plate with a built-in cleaving mechanism (Supplementary Fig.~6).
  This sample plate consists of a flag-style plate, to which a cylinder with a small arm is attached.
  The arm sits at a fixed height of 0.5 mm above the sample holder and can be rotated with a hex-key wobble stick that slots in the cylinder.
  This makes it possible to hit the ``top post'' and cleave the small and fragile samples in a controlled way.
  ARPES measurements were performed using photon energies between 20 and 180 eV, with linear horizontal and vertical light polarisations as indicated in the figure captions.
  All measurements were performed at base temperature (18 K).
  ARPES data analysis was performed using the \textit{peaks} package \cite{peaks}.

  \

  \noindent{\bf{Electronic structure calculations}}
  Density Functional Theory calculations were performed with the Quantum Espresso package \cite{QE-2009, QE-2017}.
  For bulk band structure calculations, we used fully relativistic optimized norm-conserving Vanderbilt pseudopotentials with a PBE type exchange potential \cite{PseudoDojo}, and a plane-wave cutoff energy of 84 Ry.
  Spin-orbit coupling was included.
  The unit cell parameters and atomic positions were taken from \cite{matproj} and the cell was allowed to fully relax to a total force of 2 meV/\AA \ before the band structure calculations.
  A $10\times10\times14$ $k$-mesh was used in the self-consistent field calculations.

  We obtained the bulk-projected band structures by performing bulk band structure calculations at 81 $k$-planes along the out-of-plane momentum direction ($k_{[110]}$).
  The final spectra were obtained by summing the band structures along these cuts with a small energy broadening (40 meV), and weighed with a Lorentzian factor to simulate the $k_\perp$ plane probed and corresponding $k_\perp$ broadening given a photon energy (50, 68 or 87 eV) and mean free path (here taken to be 2.5 \AA).

  For calculations of the $(110)$- and $(100)$-surface electronic structure, we performed DFT on a symmetric 12 and 10-layer slab with a stoichiometric termination on both sides.
  Scalar relativistic pseudopotentials were used with a PBE exchange potential \cite{SSSP}, and a plane-wave cutoff energy of 75 Ry on a $k$-grid of $7\times7\times1$.
  Spin-orbit coupling was not included.
  The surface layers of the slab were allowed to relax to 10 meV/\AA \ before band structure calculations.

  Additionally, we used Wannier90 to obtain a tight binding model based on the maximally localized Wannier functions \cite{Wannier90}.
  We included the Ru-4$d$ and O-2$p$ orbitals.
  This tight binding model was used for calculations of the bulk Fermi surface, and for the Surface Green's Functions calculations.
  The latter were performed within the WannierTools package \cite{WannierTools}.
  The bulk projected Fermi surfaces were obtained by taking 100 slices through this calculated bulk Fermi surface along the out-of-plane direction ($k_{[110]}$ or $k_{[100]}$), and summing these cuts with a small broadening (40 meV) and Lorentzian factor to simulate the plane that is probed at 68 eV, assuming a mean free path of 2.5 \AA.

  \section*{Acknowledgements}
  We thank Simon Moser, Luke Rhodes and Carsten Putzke for useful discussions.
  We gratefully acknowledge support from the Engineering and Physical Sciences Research Council (under Grant Nos.~EP/X015556/1 and EP/M023958/1).
  We gratefully acknowledge MAX IV Laboratory for time on the Bloch beamline under Proposal Nos. 20231118 and 20240300.
  Research conducted at MAX IV, a Swedish national user facility, is supported by the Swedish Research council under contract 2018-07152, the Swedish Governmental Agency for Innovation Systems under contract 2018-04969, and Formas under contract 2019-02496.
  For the purpose of open access, the authors have applied a Creative Commons Attribution (CC BY) licence to any Author Accepted Manuscript version arising.

  \section*{Data Availability Statement}
  The research data supporting this publication can be accessed at \url{ [[DOI TO BE INSERTED]] }.
  %
  %
}

\bibliography{references}

@article{Over:2012,
  author = {Over, Herbert},
  title = {Surface Chemistry of Ruthenium Dioxide in Heterogeneous Catalysis and
           Electrocatalysis: From Fundamental to Applied Research},
  journal = {Chemical Reviews},
  volume = {112},
  number = {6},
  pages = {3356-3426},
  year = {2012},
  doi = {10.1021/cr200247n},
  URL = {https://doi.org/10.1021/cr200247n},
}

@article{Jovic:2021,
  author = {Jovic, Vedran and Consiglio, Armando and Smith, Kevin E. and Jozwiak
            , Chris and Bostwick, Aaron and Rotenberg, Eli and Di Sante, Domenico
            and Moser, Simon},
  title = {Momentum for Catalysis: How Surface Reactions Shape the {RuO$_2$}
           Flat Surface State},
  journal = {ACS Catalysis},
  volume = {11},
  number = {3},
  pages = {1749-1757},
  year = {2021},
  doi = {10.1021/acscatal.0c04871},
  URL = {https://doi.org/10.1021/acscatal.0c04871},
}

@article{Sun:2017,
  title = {Dirac nodal lines and induced spin Hall effect in metallic rutile
           oxides},
  author = {Sun, Yan and Zhang, Yang and Liu, Chao-Xing and Felser, Claudia and
            Yan, Binghai},
  journal = {Physical Review B},
  volume = {95},
  issue = {23},
  pages = {235104},
  numpages = {7},
  year = {2017},
  month = {Jun},
  publisher = {American Physical Society},
  doi = {10.1103/PhysRevB.95.235104},
  url = {https://link.aps.org/doi/10.1103/PhysRevB.95.235104},
}

@article{Ho:2025,
  title = {Symmetry-breaking induced surface magnetization in nonmagnetic {
           RuO$_2$}},
  author = {Ho, Dai Q. and To, D. Quang and Hu, Ruiqi and Bryant, Garnett W. and
            Janotti, Anderson},
  journal = {Physical Review Materials},
  volume = {9},
  issue = {9},
  pages = {094406},
  numpages = {11},
  year = {2025},
  month = {Sep},
  publisher = {American Physical Society},
  doi = {10.1103/6fxv-153y},
  url = {https://link.aps.org/doi/10.1103/6fxv-153y},
}

@article{Jovic:2018,
  title = {Dirac nodal lines and flat-band surface state in the functional oxide
           {RuO$_2$}},
  author = {Jovic, Vedran and Koch, Roland J. and Panda, Swarup K. and Berger,
            Helmuth and Bugnon, Philippe and Magrez, Arnaud and Smith, Kevin E.
            and Biermann, Silke and Jozwiak, Chris and Bostwick, Aaron and
            Rotenberg, Eli and Moser, Simon},
  journal = {Physical Review B},
  volume = {98},
  issue = {24},
  pages = {241101},
  numpages = {5},
  year = {2018},
  month = {Dec},
  publisher = {American Physical Society},
  doi = {10.1103/PhysRevB.98.241101},
  url = {https://link.aps.org/doi/10.1103/PhysRevB.98.241101},
}

@article{Graebner:1976,
  title = {Magnetothermal oscillations in {RuO$_2$}, {OsO$_2$}, and {IrO$_2$}},
  author = {Graebner, J. E. and Greiner, E. S. and Ryden, W. D.},
  journal = {Physical Review B},
  volume = {13},
  issue = {6},
  pages = {2426--2432},
  numpages = {0},
  year = {1976},
  month = {Mar},
  publisher = {American Physical Society},
  doi = {10.1103/PhysRevB.13.2426},
  url = {https://link.aps.org/doi/10.1103/PhysRevB.13.2426},
}

@article{Wu:2025,
  title = {Fermi Surface of {RuO$_2$} Measured by Quantum Oscillations},
  author = {Wu, Zheyu and Long, Mengmeng and Chen, Hanyi and Paul, Shubhankar
            and Matsuki, Hisakazu and Zheliuk, Oleksandr and Zeitler, Uli and Li,
            Gang and Zhou, Rui and Zhu, Zengwei and Graf, Dave and Weinberger,
            Theodore I. and Grosche, F. Malte and Maeno, Yoshiteru and Eaton,
            Alexander G.},
  journal = {Physical Review X},
  volume = {15},
  issue = {3},
  pages = {031044},
  numpages = {13},
  year = {2025},
  month = {Aug},
  publisher = {American Physical Society},
  doi = {10.1103/5js8-2hj8},
  url = {https://link.aps.org/doi/10.1103/5js8-2hj8},
}

@article{Liu:2024,
  title = {Absence of Altermagnetic Spin Splitting Character in Rutile Oxide {
           RuO$_2$}},
  author = {Liu, Jiayu and Zhan, Jie and Li, Tongrui and Liu, Jishan and Cheng,
            Shufan and Shi, Yuming and Deng, Liwei and Zhang, Meng and Li, Chihao
            and Ding, Jianyang and Jiang, Qi and Ye, Mao and Liu, Zhengtai and
            Jiang, Zhicheng and Wang, Siyu and Li, Qian and Xie, Yanwu and Wang,
            Yilin and Qiao, Shan and Wen, Jinsheng and Sun, Yan and Shen, Dawei},
  journal = {Physical Review Letters},
  volume = {133},
  issue = {17},
  pages = {176401},
  numpages = {7},
  year = {2024},
  month = {Oct},
  publisher = {American Physical Society},
  doi = {10.1103/PhysRevLett.133.176401},
  url = {https://link.aps.org/doi/10.1103/PhysRevLett.133.176401},
}

@article{Osumi:2026,
  title = {Spin-degenerate bulk bands and topological surface states associated
           with Dirac nodal lines in {RuO$_2$}},
  author = {Osumi, Takumi and Yamauchi, Kunihiko and Souma, Seigo and Paul,
            Shubhankar and Honma, Asuka and Nakayama, Kosuke and Ozawa, Kenichi
            and Kitamura, Miho and Horiba, Koji and Kumigashira, Hiroshi and Bigi
            , Chiara and Bertran, Fran\c{c}ois and Oguchi, Tamio and Takahashi,
            Takashi and Maeno, Yoshiteru and Sato, Takafumi},
  journal = {Physical Review B},
  volume = {113},
  issue = {8},
  pages = {085116},
  numpages = {18},
  year = {2026},
  month = {Feb},
  publisher = {American Physical Society},
  doi = {10.1103/wvs6-hqfv},
  url = {https://link.aps.org/doi/10.1103/wvs6-hqfv},
}

@article{Hunter:2024,
  title = {Controlling crystal cleavage in focused ion beam shaped specimens for
           surface spectroscopy},
  author = {Hunter, A. and Putzke, C. and Gaponenko, I. and Tamai, A. and
            Baumberger, F. and Moll, P. J. W.},
  journal = {Review of Scientific Instruments},
  volume = {95},
  issue = {3},
  pages = {033905},
  year = {2024},
  month = {March},
  doi = {10.1063/5.0186480},
  url = {
         https://pubs.aip.org/rsi/article/95/3/033905/3270229/Controlling-crystal-cleavage-in-focused-ion-beam
         },
}

@article{QE-2009,
  doi = {10.1088/0953-8984/21/39/395502},
  url = {https://doi.org/10.1088/0953-8984/21/39/395502},
  year = {2009},
  month = {sep},
  publisher = {},
  volume = {21},
  number = {39},
  pages = {395502},
  author = {Giannozzi, Paolo and Baroni, Stefano and Bonini, Nicola and Calandra
            , Matteo and Car, Roberto and Cavazzoni, Carlo and Ceresoli, Davide
            and Chiarotti, Guido L and Cococcioni, Matteo and Dabo, Ismaila and
            Dal Corso, Andrea and de Gironcoli, Stefano and Fabris, Stefano and
            Fratesi, Guido and Gebauer, Ralph and Gerstmann, Uwe and Gougoussis,
            Christos and Kokalj, Anton and Lazzeri, Michele and Martin-Samos,
            Layla and Marzari, Nicola and Mauri, Francesco and Mazzarello,
            Riccardo and Paolini, Stefano and Pasquarello, Alfredo and Paulatto,
            Lorenzo and Sbraccia, Carlo and Scandolo, Sandro and Sclauzero,
            Gabriele and Seitsonen, Ari P and Smogunov, Alexander and Umari,
            Paolo and Wentzcovitch, Renata M},
  title = {{QUANTUM ESPRESSO}: a modular and open-source software project for
           quantum simulations of materials},
  journal = {Journal of Physics: Condensed Matter},
}

@article{QE-2017,
  doi = {10.1088/1361-648X/aa8f79},
  url = {https://doi.org/10.1088/1361-648X/aa8f79},
  year = {2017},
  month = {oct},
  publisher = {IOP Publishing},
  volume = {29},
  number = {46},
  pages = {465901},
  author = {Giannozzi, P and Andreussi, O and Brumme, T and Bunau, O and
            Buongiorno Nardelli, M and Calandra, M and Car, R and Cavazzoni, C
            and Ceresoli, D and Cococcioni, M and Colonna, N and Carnimeo, I and
            Dal Corso, A and de Gironcoli, S and Delugas, P and DiStasio, R A and
            Ferretti, A and Floris, A and Fratesi, G and Fugallo, G and Gebauer,
            R and Gerstmann, U and Giustino, F and Gorni, T and Jia, J and
            Kawamura, M and Ko, H-Y and Kokalj, A and {K\"{u}\c{c}\"{u}kbenli}, E
            and Lazzeri, M and Marsili, M and Marzari, N and Mauri, F and Nguyen,
            N L and Nguyen, H-V and Otero-de-la-Roza, A and Paulatto, L and Ponc
            \'e, S and Rocca, D and Sabatini, R and Santra, B and Schlipf, M and
            Seitsonen, A P and Smogunov, A and Timrov, I and Thonhauser, T and
            Umari, P and Vast, N and Wu, X and Baroni, S},
  title = {Advanced capabilities for materials modelling with {QUANTUM ESPRESSO}
           },
  journal = {Journal of Physics: Condensed Matter},
}

@article{Wannier90,
  doi = {10.1088/1361-648X/ab51ff},
  url = {https://doi.org/10.1088/1361-648X/ab51ff},
  year = {2020},
  month = {jan},
  publisher = {IOP Publishing},
  volume = {32},
  number = {16},
  pages = {165902},
  author = {Pizzi, Giovanni and Vitale, Valerio and Arita, Ryotaro and Bl\"ugel,
            Stefan and Freimuth, Frank and G\'eranton, Guillaume and Gibertini,
            Marco and Gresch, Dominik and Johnson, Charles and Koretsune, Takashi
            and {Iba\~nez-Azpiroz}, Julen and Lee, Hyungjun and Lihm, Jae-Mo and
            Marchand, Daniel and Marrazzo, Antimo and Mokrousov, Yuriy and
            Mustafa, Jamal I and Nohara, Yoshiro and Nomura, Yusuke and Paulatto,
            Lorenzo and Ponc\'e, Samuel and Ponweiser, Thomas and Qiao, Junfeng
            and Thöle, Florian and Tsirkin, Stepan S and Wierzbowska, Ma\l{}
            gorzata and Marzari, Nicola and Vanderbilt, David and Souza, Ivo and
            Mostofi, Arash A and Yates, Jonathan R},
  title = {Wannier90 as a community code: new features and applications},
  journal = {Journal of Physics: Condensed Matter},
}

@article{WannierTools,
  title = {WannierTools: An open-source software package for novel topological
           materials},
  journal = {Computer Physics Communications},
  volume = {224},
  pages = {405-416},
  year = {2018},
  issn = {0010-4655},
  doi = {https://doi.org/10.1016/j.cpc.2017.09.033},
  url = {https://www.sciencedirect.com/science/article/pii/S0010465517303442},
  author = {QuanSheng Wu and ShengNan Zhang and Hai-Feng Song and Matthias
            Troyer and Alexey A. Soluyanov},
}

@article{PseudoDojo,
  title = {Optimized norm-conserving Vanderbilt pseudopotentials},
  author = {Hamann, D. R.},
  journal = {Phys. Rev. B},
  volume = {88},
  issue = {8},
  pages = {085117},
  numpages = {10},
  year = {2013},
  month = {Aug},
  publisher = {American Physical Society},
  doi = {10.1103/PhysRevB.88.085117},
  url = {https://link.aps.org/doi/10.1103/PhysRevB.88.085117},
}

@article{SSSP,
  title = {Precision and efficiency in solid-state pseudopotential calculations},
  author = {Prandini, Gianluca and Marrazzo, Antimo and Castelli, Ivano E and
            Mounet, Nicolas and Marzari, Nicola},
  journal = {npj Computational Materials},
  volume = {4},
  number = {1},
  pages = {72},
  year = {2018},
  issn = {2057-3960},
  url = {https://www.nature.com/articles/s41524-018-0127-2},
  doi = {10.1038/s41524-018-0127-2},
  publisher = {Nature Publishing Group UK London},
}

@article{Smejkal:2022,
  title = {Beyond Conventional Ferromagnetism and Antiferromagnetism: A Phase
           with Nonrelativistic Spin and Crystal Rotation Symmetry},
  author = {\ifmmode \check{S}\else \v{S}\fi{}mejkal, Libor and Sinova, Jairo
            and Jungwirth, Tomas},
  journal = {Physical Review X},
  volume = {12},
  issue = {3},
  pages = {031042},
  numpages = {16},
  year = {2022},
  month = {Sep},
  publisher = {American Physical Society},
  doi = {10.1103/PhysRevX.12.031042},
  url = {https://link.aps.org/doi/10.1103/PhysRevX.12.031042},
}

@article{Hiraishi:2024,
  title = {Nonmagnetic Ground State in {RuO$_2$} Revealed by Muon Spin Rotation},
  author = {Hiraishi, M. and Okabe, H. and Koda, A. and Kadono, R. and Muroi, T.
            and Hirai, D. and Hiroi, Z.},
  journal = {Physical Review Letters},
  volume = {132},
  issue = {16},
  pages = {166702},
  numpages = {6},
  year = {2024},
  month = {Apr},
  publisher = {American Physical Society},
  doi = {10.1103/PhysRevLett.132.166702},
  url = {https://link.aps.org/doi/10.1103/PhysRevLett.132.166702},
}

@article{Kessler:2024,
  title = {Absence of magnetic order in {RuO2}: insights from $\mu${SR}
           spectroscopy and neutron diffraction},
  volume = {2},
  issn = {2948-2119},
  url = {https://doi.org/10.1038/s44306-024-00055-y},
  doi = {10.1038/s44306-024-00055-y},
  number = {1},
  journal = {npj Spintronics},
  author = {{Ke{\ss}ler}, Philipp and Garcia-Gassull, Laura and Suter, Andreas
            and Prokscha, Thomas and Salman, Zaher and Khalyavin, Dmitry and
            Manuel, Pascal and Orlandi, Fabio and Mazin, Igor I. and {Valent\'i},
            Roser and Moser, Simon},
  month = oct,
  year = {2024},
  pages = {50},
}

@article{Torun:2013,
  author = {Torun, E. and Fang, C. M. and {de Wijs}, G. A. and {de Groot}, R. A.
            },
  title = {Role of Magnetism in Catalysis: {RuO$_2$} (110) Surface},
  journal = {The Journal of Physical Chemistry C},
  volume = {117},
  number = {12},
  pages = {6353-6357},
  year = {2013},
  doi = {10.1021/jp4020367},
  URL = {https://doi.org/10.1021/jp4020367},
}

@article{Gonzalez:2021,
  title = {Efficient Electrical Spin Splitter Based on Nonrelativistic Collinear
           Antiferromagnetism},
  author = {Gonz\'alez-Hern\'andez, Rafael and \v{S}mejkal, Libor and V\'yborn\'
            y, Karel and Yahagi, Yuta and Sinova, Jairo and Jungwirth, Tom\'{a}\v
            {s} and \v{Z}elezn\'{y}, Jakub},
  journal = {Physical Review Letters},
  volume = {126},
  issue = {12},
  pages = {127701},
  numpages = {6},
  year = {2021},
  month = {Mar},
  publisher = {American Physical Society},
  doi = {10.1103/PhysRevLett.126.127701},
  url = {https://link.aps.org/doi/10.1103/PhysRevLett.126.127701},
}

@article{Feng:2022,
  title = {An anomalous {Hall} effect in altermagnetic ruthenium dioxide},
  volume = {5},
  issn = {2520-1131},
  url = {https://doi.org/10.1038/s41928-022-00866-z},
  doi = {10.1038/s41928-022-00866-z},
  number = {11},
  journal = {Nature Electronics},
  author = {Feng, Zexin and Zhou, Xiaorong and Šmejkal, Libor and Wu, Lei and
            Zhu, Zengwei and Guo, Huixin and González-Hernández, Rafael and Wang,
            Xiaoning and Yan, Han and Qin, Peixin and Zhang, Xin and Wu, Haojiang
            and Chen, Hongyu and Meng, Ziang and Liu, Li and Xia, Zhengcai and
            Sinova, Jairo and Jungwirth, Tomáš and Liu, Zhiqi},
  month = nov,
  year = {2022},
  pages = {735--743},
}

@article{Tschirner:2023,
  title = {Saturation of the anomalous {Hall} effect at high magnetic fields in
           altermagnetic {RuO$_2$}},
  volume = {11},
  issn = {2166-532X},
  url = {https://doi.org/10.1063/5.0160335},
  doi = {10.1063/5.0160335},
  number = {10},
  journal = {APL Materials},
  author = {Tschirner, Teresa and Keßler, Philipp and Gonzalez Betancourt, Ruben
            Dario and Kotte, Tommy and Kriegner, Dominik and Büchner, Bernd and
            Dufouleur, Joseph and Kamp, Martin and Jovic, Vedran and Smejkal,
            Libor and Sinova, Jairo and Claessen, Ralph and Jungwirth, Tomas and
            Moser, Simon and Reichlova, Helena and Veyrat, Louis},
  month = oct,
  year = {2023},
  pages = {101103},
}

@article{Bai:2023,
  title = {Efficient Spin-to-Charge Conversion via Altermagnetic Spin Splitting
           Effect in Antiferromagnet {RuO$_2$}},
  author = {Bai, H. and Zhang, Y. C. and Zhou, Y. J. and Chen, P. and Wan, C. H.
            and Han, L. and Zhu, W. X. and Liang, S. X. and Su, Y. C. and Han, X.
            F. and Pan, F. and Song, C.},
  journal = {Physical Review Letters},
  volume = {130},
  issue = {21},
  pages = {216701},
  numpages = {6},
  year = {2023},
  month = {May},
  publisher = {American Physical Society},
  doi = {10.1103/PhysRevLett.130.216701},
  url = {https://link.aps.org/doi/10.1103/PhysRevLett.130.216701},
}

@article{Karube:2022,
  title = {Observation of Spin-Splitter Torque in Collinear Antiferromagnetic {
           RuO$_2$}},
  author = {Karube, Shutaro and Tanaka, Takahiro and Sugawara, Daichi and
            Kadoguchi, Naohiro and Kohda, Makoto and Nitta, Junsaku},
  journal = {Physical Review Letters},
  volume = {129},
  issue = {13},
  pages = {137201},
  numpages = {6},
  year = {2022},
  month = {Sep},
  publisher = {American Physical Society},
  doi = {10.1103/PhysRevLett.129.137201},
  url = {https://link.aps.org/doi/10.1103/PhysRevLett.129.137201},
}

@article{Kiefer:2025,
  doi = {10.1088/1361-648X/adad2a},
  url = {https://doi.org/10.1088/1361-648X/adad2a},
  year = {2025},
  month = {feb},
  publisher = {IOP Publishing},
  volume = {37},
  number = {13},
  pages = {135801},
  author = {Kiefer, L and Wirth, F and Bertin, A and Becker, P and Bohat\'y, L
            and Schmalzl, K and Stunault, A and Rodr\'iguez-Velamazan, J A and
            Fabelo, O and Braden, M},
  title = {Crystal structure and absence of magnetic order in single-crystalline
           {RuO$_2$}},
  journal = {Journal of Physics: Condensed Matter},
}

@article{Lin:2025,
  title = {Bulk band structure of {RuO$_2$} measured with soft x-ray
           angle-resolved photoemission spectroscopy},
  author = {Lin, Zihan and Chen, Dong and Lu, Wenlong and Liang, Xin and Feng,
            Shiyu and Yamagami, Kohei and Osiecki, Jacek and Leandersson, Mats
            and Thiagarajan, Balasubramanian and Liu, Junwei and Felser, Claudia
            and Ma, Junzhang},
  journal = {Physical Review B},
  volume = {111},
  issue = {13},
  pages = {134450},
  numpages = {9},
  year = {2025},
  month = {Apr},
  publisher = {American Physical Society},
  doi = {10.1103/PhysRevB.111.134450},
  url = {https://link.aps.org/doi/10.1103/PhysRevB.111.134450},
}

@article{peaks,
  title = {peaks: a {Python} package for analysis of angle-resolved
           photoemission and related spectroscopies},
  author = {Phil D. C. King and Brendan Edwards and Shu Mo and Tommaso Antonelli
            and Edgar Abarca Morales and Lewis Hart and Liam Trzaska},
  year = {2025},
  journal = {{arXiv}},
  volume = {preprint arXiv:2508.04803},
  url = {https://arxiv.org/abs/2508.04803},
}

@article{Yavorsky:1996,
  title = {Ab initio calculation of the Fermi surface of {RuO$_2$}},
  journal = {Physica B: Condensed Matter},
  volume = {225},
  number = {3},
  pages = {243-250},
  year = {1996},
  issn = {0921-4526},
  doi = {https://doi.org/10.1016/0921-4526(96)00270-0},
  url = {https://www.sciencedirect.com/science/article/pii/0921452696002700},
  author = {B.Yu. Yavorsky and O.V. Krasovska and E.E. Krasovskii and A.N.
            Yaresko and V.N. Antonov},
}

@article{matproj,
  title = {Materials Data on {RuO$_2$} by Materials Project},
  abstractNote = {RuO2 is Hydrophilite-like structured and crystallizes in the
                  tetragonal P4_2/mnm space group. The structure is
                  three-dimensional. Ru4+ is bonded to six equivalent O2- atoms
                  to form a mixture of edge and corner-sharing RuO6 octahedra.
                  The corner-sharing octahedral tilt angles are 52°. There is two
                  shorter (1.96 Å) and four longer (2.01 Å) Ru–O bond length. O2-
                  is bonded in a trigonal planar geometry to three equivalent
                  Ru4+ atoms.},
  doi = {10.17188/1307989},
  journal = {},
  number = {},
  volume = {},
  place = {United States},
  year = {2020},
  month = {5},
}

@article{oppermann1975,
  title = {Zum chemischen Transport der Übergangsmetalldioxide mit
           Tellurhalogeniden},
  author = {Oppermann, H. and Ritschel, M.},
  year = {1975},
  journal = {Kristall und Technik},
  shortjournal = {Krist. Techn.},
  volume = {10},
  number = {5},
  pages = {485--503},
  issn = {00234753, 15214079},
  doi = {10.1002/crat.19750100504},
  url = {https://onlinelibrary.wiley.com/doi/10.1002/crat.19750100504},
  urldate = {2023-02-20},
  langid = {ngerman},
}

@article{wadehra_strain-induced_2025,
  title = {Strain-induced superconductivity in {RuO$_2$}$(100)$ thin-films},
  volume = {6},
  issn = {2662-4443},
  url = {https://www.nature.com/articles/s43246-025-00856-6},
  doi = {10.1038/s43246-025-00856-6},
  number = {1},
  urldate = {2026-03-30},
  journal = {Communications Materials},
  author = {Wadehra, Neha and Gregory, Benjamin Z. and Zhang, Shuyuan and
            Schnitzer, Noah and Iguchi, Yusuke and Li, Yilin Evan and Pamuk,
            Betül and Muller, David A. and Singer, Andrej and Shen, Kyle M. and
            Schlom, Darrell G.},
  month = jul,
  year = {2025},
  pages = {135},
}

@article{Ruf:2021,
  title = {Strain-stabilized superconductivity},
  volume = {12},
  copyright = {2021 The Author(s)},
  issn = {2041-1723},
  url = {https://www.nature.com/articles/s41467-020-20252-7},
  doi = {10.1038/s41467-020-20252-7},
  number = {1},
  urldate = {2024-08-20},
  journal = {Nature Communications},
  publisher = {Nature Publishing Group},
  author = {Ruf, J. P. and Paik, H. and Schreiber, N. J. and Nair, H. P. and
            Miao, L. and Kawasaki, J. K. and Nelson, J. N. and Faeth, B. D. and
            Lee, Y. and Goodge, B. H. and Pamuk, B. and Fennie, C. J. and
            Kourkoutis, L. F. and Schlom, D. G. and Shen, K. M.},
  month = jan,
  year = {2021},
  pages = {59},
}

@article{Over:2002,
  title = {Experimental and simulated {STM} images of stoichiometric and
           partially reduced {RuO$_2$}(110) surfaces including adsorbates},
  journal = {Surface Science},
  volume = {515},
  number = {1},
  pages = {143-156},
  year = {2002},
  issn = {0039-6028},
  doi = {https://doi.org/10.1016/S0039-6028(02)01853-8},
  url = {https://www.sciencedirect.com/science/article/pii/S0039602802018538},
  author = {H. Over and A.P. Seitsonen and E. Lundgren and M. Schmid and P.
            Varga},
}

\ifarXiv
\foreach \x in {1,...,\numbersupplementpages}
{
  \clearpage
  \includepdf[pages={\x}]{\supplementfilename}
}
\fi

\end{document}